\documentclass{aa}
\usepackage{graphicx}
\usepackage[varg]{txfonts}
\usepackage{natbib}
\usepackage{float}
\usepackage{color}

\begin{document}
  \title{Violent mergers of nearly equal-mass white dwarf as progenitors of subluminous Type Ia supernovae}

  \author{R.~Pakmor$^{1,2}$ \and
          S.~Hachinger$^1$ \and
          F.~K.~R\"{o}pke$^{1,3}$ \and
          W.~Hillebrandt$^1$
          }

  \institute{$^1$Max-Planck-Institut f\"{u}r Astrophysik, Karl-Schwarzschild-Str. 1, 85741 Garching, Germany \\
             $^2$Heidelberger Institut f\"{u}r Theoretische Studien, Schloss-Wolfsbrunnenweg 35, 69118 Heidelberg, Germany\\
             $^3$Institut f{\"u}r Theoretische Physik und Astrophysik, Universit{\"a}t W{\"u}rzburg, Am Hubland, 97074 W{\"u}rzburg, Germany\\
          \email{rpakmor@mpa-garching.mpg.de}
          }

  \authorrunning{R.~Pakmor et al.}
  \titlerunning{Violent mergers of nearly equal-mass white dwarf}

  \date{Received ; accepted }

  \abstract
  {The origin of subluminous Type Ia supernovae (SNe Ia) has long eluded
   any explanation, as all Chandrasekhar-mass models have severe
   problems reproducing them. Recently, it has been proposed that
   violent mergers of two white dwarfs of $0.9 \mathrm{M_\odot}$
   could lead to subluminous SNe Ia events that resemble
     1991bg-like SNe~Ia.}
  {Here we investigate whether this scenario still works for mergers
   of two white dwarfs with a mass ratio smaller than one. We aim to determine
   the range of mass ratios for which a detonation still forms during
   the merger, as only those events will lead to a SN Ia. This range is an
   important ingredient for population synthesis and one decisive point
   to judge the viability of the scenario. In addition,
   we perform a resolution study of one of the models. Finally we discuss the connection
   between violent white dwarf mergers with a primary mass of $0.9 \mathrm{M_\odot}$
   and 1991bg-like SNe Ia.}
  {The latest version of the smoothed particle hydrodynamics code \textsc{Gadget3} is
    used to evolve binary systems with different mass ratios until they merge. We analyze
    the result and look for hot spots in which detonations can form.}
  {We show that mergers of two white dwarfs with a primary white dwarf mass of $\approx 0.9 \, \mathrm{M_\odot}$
    and a mass ratio larger than about $0.8$ robustly reach the conditions we require to
    ignite a detonation and thus produce thermonuclear explosions during the merger itself.
    We also find that while our simulations do not yet completely resolve the hot spots, 
    increasing the resolution leads to conditions that are even more likely to ignite
    detonations.
    Additionally we compare the abundance structure of the ejecta of the thermonuclear 
    explosion of two merged white dwarfs with data inferred from observations of a 1991bg-like 
    SN Ia (SN 2005bl). The abundance distributions of intermediate mass and iron group elements 
    in velocity space agree qualitatively, and our model reproduces the lack of material at
    high velocities inferred from the observations.}
  {The violent merger scenario constitutes a robust possibility for two merging white dwarfs
    to produce a thermonuclear explosion.
    Mergers with a primary white dwarf mass of $\approx 0.9 \, \mathrm{M_\odot}$ are
    very promising candidates for explaining subluminous SNe Ia. This would imply that subluminous SNe~Ia
    form a distinct class of objects, that are not produced in the standard single white
    dwarf scenario for SNe Ia, but instead arise from a different
    progenitor channel and explosion mechanism.}


  \maketitle

  \section{Introduction}
  \label{sec:introduction}

  Type Ia supernovae are among the brightest objects in the
  Universe. Their apparent homogeneity makes them one of the most
  important probes for cosmic expansion. This
    application relies on an empirical luminosity calibration
  \citep[e.g.][]{phillips1993a}. Only recently this calibration has been justified
  by theoretical studies, as models were able to reproduce it in terms
  of the underlying physics \citep{kasen2009a, mazzali2001a}. The
  standard SNe Ia \citep{branch1993a} used as cosmological distance indicators make up about $70\%$ \citep{li2010a}
  of the observed SNe Ia. There is, however, a number of subclasses of
  SNe Ia consisting of very dim (e.g. SN~1991bg, \citealt{leibundgut1993a}), normal, bright
  (e.g. SN~1991T, \citealt{filippenko1992a}) and very bright supernovae (e.g. SN~2007if, \citealt{scalzo2010a}).
  Until today, neither the progenitor scenario nor the
  explosion mechanism of SNe Ia are completely understood.
  In particular, very dim and very bright objects are
  difficult to explain. There is general consensus only
  that all SNe Ia are thermonuclear explosions of white dwarfs (WDs) in binary
  systems. The nature of the companion star and the details of the
  explosion mechanism, however, are still debated.

  Current progenitor models distinguish between the \emph{single
    degenerate} and the \emph{double degenerate} scenario. The former
  \citep[proposed by][]{whelan1973a} assumes a main sequence or giant
  star, i.e.~a non-degenerate star, to be the companion of the
  exploding white dwarf. It feeds the white dwarf via Roche-lobe
  overflow or by winds until the white dwarf approaches the
  Chandrasekhar mass and explodes. In contrast, the double degenerate scenario
  \citep{iben1984a, webbink1984a} assumes a binary system of
  two carbon-oxygen white dwarfs. Despite being favored by stellar population
  synthesis studies \citep[e.g.][]{ruiter2009a}, this scenario has
  received little attention \citep[see,
  e.g.][]{hillebrandt2000a}. \citet{benz1990a} found that in a binary
  system of a $1.2 \,\mathrm{M_\odot}$ and a $0.9 \, \mathrm{M_\odot}$
  white dwarf the less massive white dwarf is disrupted and accreted
  onto the more massive white dwarf. This happens within a few orbits after mass
  transfer between the stars has started. It has also been shown
  \citep[e.g.][]{nomoto1985a,saio1998a} that the most likely result
  of such a merger is an inward propagating deflagration flame that
  converts the white dwarf into an O-Ne white dwarf. This white dwarf
  then undergoes accretion-induced collapse rather than
  thermonuclear explosion \citep{nomoto1991a}. Recent
  high-resolution simulations of white dwarf mergers with careful
  treatment of the initial conditions \citep[][]{dsouza2006a,
    motl2007a, dan2008a}, however, show indications that the immediate
  disruption of the less massive white dwarf may be avoided. Instead,
  they find stable mass transfer for more than $50$ orbits. When
  the less massive secondary WD is disrupted after these orbits, the
  accretion onto the primary is slower and carbon shell ignition may be
  avoided \citep{yoon2007a}.

  In the Chandrasekhar-mass scenario, a white dwarf is thought to ignite near
  the center. At first the flame propagates
  subsonically as a deflagration, but burns only a fraction of the
  star. The energy release expands the star and in a later stage of the explosion,
  a detonation triggers that propagates supersonically. The mechanism that leads to
  the formation of the detonation is still an open question. One of
  the most promising scenarios is a deflagration to detonation
  transition by turbulence \citep[e.g.][]{khokhlov1997a,roepke2007b,roepke2007c}.
  Other proposed mechanisms are the gravitationally confined
  detonation \citep[e.g.][]{jordan2008a,meakin2009a} and pulsating
  reverse detonation \citep{bravo2006a}. Note that a thermonuclear explosion 
  may also be possible before the white dwarf reaches the Chandrasekhar-mass if
  the carbon-oxygen white dwarf accretes helium from a non-degenerate or white dwarf
  companion \citep{fink2007a,fink2010a,sim2010a,guillochon2010a}.

  A more detailed comparison of different explosion mechanisms can be found in
  \citet{hillebrandt2000a}. However, none of the Chandrasekhar-mass models are able to
  explain very dim SN Ia with a $^{56}\mathrm{Ni}$ mass of $0.2 \,
  \mathrm{M_\odot}$ or less.

  Recently, \citet{pakmor2010a} proposed that these SNe Ia
  are the result of a violent merger of two equally massive white dwarfs.
  Once a binary system of two white dwarfs has formed, gravitational
  wave emission leads to continuous shrinking of its orbit. At some
  point, the two white dwarfs are close enough to interact
  directly. If the mass ratio of the two white dwarfs is close to
  unity, the interaction becomes very violent and the slightly
  less massive, and therefore less compact white dwarf is destroyed.
  Its material plunges onto the remaining white dwarf,
  is compressed and heats up. At the interface between the merging WDs 
  material can become hot enough for carbon to ignite. As carbon burning starts,
  it is heated up further. If these hot spots occur at high enough densities,
  they can reach the conditions that are necessary to ignite a detonation.

  The detonation then propagates through the merged object and burns most of
  the original carbon-oxygen material to heavier nuclei. As the detonation flame
  moves supersonically with respect to the fuel and the timescale of
  the flame to cross the whole star is much shorter than dynamical
  timescales involved, the system is burned
  nearly instantaneously at the moment the detonation is ignited.
  With enough energy released from nuclear burning to overcome the
  gravitational binding energy, the merged object explodes, and
  finally reaches a state of homologous expansion.

  In this work, we explore mergers of white dwarfs with a primary mass of
  $0.9 \,\mathrm{M_\odot}$ and mass ratios between $0.8$ and one. We investigate
  how the conditions during the merging event change with mass ratio and determine
  for which range of mass ratios the merger will produce a thermonuclear
  explosion.

  The paper is structured as follows.
  In Section \ref{sec:implementation}, the SPH code used in this work is
  summarized and the modifications are presented that are introduced to
  adapt the code to the problem. The initial setup is described in Section \ref{sec:setup}.
  Section \ref{sec:merger} investigates inspiral and merger of one specific binary system
  and discusses the conditions that lead to the ignition of a detonation. Section \ref{sec:mass_ratio}
  compares mergers with different mass ratios. In Section \ref{sec:observations} we compare the
  outcome of a thermonuclear explosion following a violent merger with observations.
  Conclusions are drawn in Section \ref{sec:conclusion}.

  \section{Implementation}
  \label{sec:implementation}
  
  In order to simulate the inspiral of the binary system, the latest version 
  of the smoothed particle hydrodynamics (SPH) code \textsc{Gadget}
  \citep{springel2005a} is used with some modifications. These are needed as
  \textsc{Gadget} was originally written for cosmological applications. The most important
  modifications to the code are the implementation of a stellar
  equation of state and a nuclear reaction network. 
  For the equation of state (EoS) of the electron gas, positrons and radiation we
  use a tabulated version of the EoS by \citet{timmes1999a}, in which we
  interpolate linearly. The ions are treated as a fully ionized ideal gas. 

  We implemented a 13 isotope nuclear reaction network. The network
  contains all $\mathrm{\alpha}$-elements up to $\mathrm{^{56}Ni}$ (
  $^{4}\mathrm{He}$, $^{12}\mathrm{C}$, $^{16}\mathrm{O}$,
  $^{20}\mathrm{Ne}$, $^{24}\mathrm{Mg}$, $^{28}\mathrm{Si}$,
  $^{32}\mathrm{S}$, $^{36}\mathrm{Ar}$, $^{40}\mathrm{Ca}$,
  $^{44}\mathrm{Ti}$, $^{48}\mathrm{Cr}$, $^{52}\mathrm{Fe}$,
  $^{56}\mathrm{Ni}$ ). The nuclear reaction rates are taken from the
  latest (2009) release of the REACLIB database
  \citep{rauscher2000a}. The network is solved in each timestep for
  all active particles whose temperatures exceed $10^6 \mathrm{K}$.
  It is solved separately from the hydrodynamical equations employing
  operator splitting. Typically the network requires significantly
  smaller timesteps than the hydrodynamical timesteps; therefore multiple network timesteps have to be
  performed during one hydrodynamical timestep. The change in composition
  is taken into account for the EoS. The energy release (or
  consumption) due to changes in the composition is added to the
  internal energy of a particle.
  As a consequence of the implementation of the stellar equation of state and the nuclear
  network we solve the energy equation instead of the entropy
  equation (as in the original version of \textsc{Gadget}).
  Details and tests of these modifications will be
  discussed in \citet{pakmor2011a}.

  The configuration of the \textsc{Gadget} code is very similar to its previous use in
  \citet{pakmor2008a}. All particles have the same mass. The smoothing
  length is chosen such that each particle has $50$ neighbor particles.
  To calculate the gravitational forces only the tree method is used.
  The gravitational softening length of a particle is set equal to its smoothing length.

  \begin{figure*}[!t]
     \centering
     \includegraphics[width=0.9\linewidth]{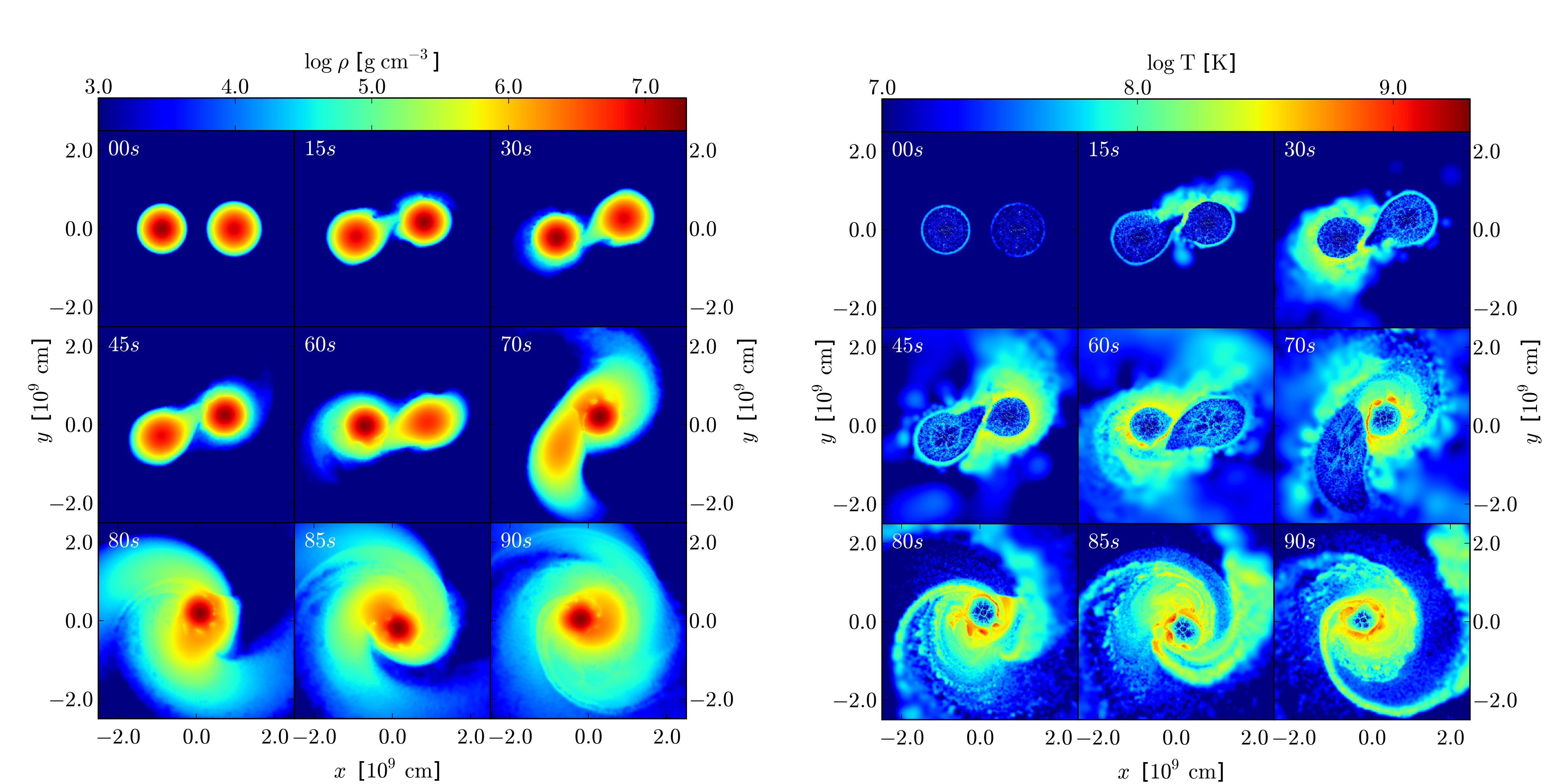}
     \caption{Snapshots of the evolution of the binary system during
       the inspiral. The system contains two white dwarfs of
       $0.90 \, \mathrm{M}_{\odot}$ and $0.81 \, \mathrm{M}_{\odot}$, respectively. It
       was set up with an initial orbital period of $33 \mathrm{s}$, rotating clockwise.
       Color-coded is the logarithm of the density on the
       left and of the temperature on the right.} 
     \label{fig:inspiral}
  \end{figure*}

  \section{Simulation setup}
  \label{sec:setup}

  In a first step, a one-dimensional model of a carbon-oxygen white dwarf
  in hydrostatic equilibrium is constructed. This is done by
  integrating the equation of hydrostatic equilibrium over
  the mass profile of a white dwarf, starting at the center with a given
  central density and assuming a constant temperature and nuclear
  composition throughout the whole star. The central density is chosen
  to give the desired mass of the white dwarf.

  All white dwarfs are assumed to have a constant temperature of
  $\mathit{T} = 5 \times 10^5 \, \mathrm{K}$ and a uniform
  composition of $\mathit{X}_\mathrm{C} = \mathit{X}_\mathrm{O} = 0.5$
  is used. To set up the binaries we create white dwarfs of 
  $\mathit{M} = 0.90 \, \mathrm{M}_\odot$, $\mathit{M} = 0.70 \, \mathrm{M}_\odot$, $\mathit{M} = 0.76 \, \mathrm{M}_\odot$,
  $\mathit{M} = 0.81 \, \mathrm{M}_\odot$ and $\mathit{M} = 0.89 \, \mathrm{M}_\odot$, respectively.

  In a second step, each one-dimensional white dwarf model is mapped to a
  distribution of SPH particles resembling its one-dimensional
  profile. Details of this procedure will be described in \citet{pakmor2011a}.
  All particles of all white dwarfs represent the same mass. The particle mass is chosen such that
  the most massive white dwarf ($\mathit{M} = 0.90 \, \mathrm{M}_\odot$) consists
  of $10^6$ particles. Using the \textsc{Gadget} code, every model is finally relaxed
  for $100 \, \mathrm{s}$ to damp out spurious numerical noise introduced by the setup.

  After relaxation, the white dwarf models are used to construct binaries. Both stars in such a binary are
  set onto a circular orbit around their combined center
  of mass. All particles of each star are given the same initial
  velocity. The period is chosen to provide a marginally stable
  orbit, i.e.~the system is stable for more than one complete orbit before tidal
  interactions lead to dynamical interactions between the white
  dwarfs.

  Note that this setup does not relax the binary system once the white dwarfs
  feel the tidal forces of their companion. Performing such an additional relaxation step may lead to
  more stable initial conditions and therefore a less violent merger, but to investigate this question
  would be computationally very demanding.

  \section{Inspiral and merger of a $0.90\mathrm{M}_{\odot}$ + $0.81\mathrm{M}_{\odot}$ white dwarf binary}
  \label{sec:merger}

  The binary evolution in the inspiral and merger phase is
  illustrated in Fig.~\ref{fig:inspiral} for the example of a merger
  between a $0.90 \, \mathrm{M}_{\odot}$ and a $0.81 \, \mathrm{M}_{\odot}$
  white dwarf. Shown are snapshots of density and temperature in slices
  through the orbital plane ($z = 0$).
  The binary system rotates clockwise. It is stable for about two orbits, but both stars
  are slightly distorted by tidal forces. In the beginning, only the less massive
  white dwarf is distorted, while the more massive star stays
  relatively unaffected for much longer.

  Mass transfer sets in a few seconds after the start of the simulation.
  After about two orbits ($\approx 60 \, \mathrm{s}$) the secondary
  white dwarf starts to break up and its material is accreted rapidly onto
  the primary white dwarf. In this process, material at the surface of
  the primary white dwarf is strongly heated up, first by compression and
  later also by nuclear reactions to temperatures well above $10^9\, \mathrm{K}$.
  This process starts with compression that subsequently leads to the
  ignition of carbon burning. The temperature increase beyond $1.5 \times 10^9\, \mathrm{K}$
  is due to the energy release in this nuclear reaction.

  The evolution of the binary is qualitatively similar to previous
  studies of merging double degenerate systems
  \citep[e.g.][]{guerrero2004a, yoon2007a, loren2009a}. In these
  studies, the less massive WD is disrupted in the merger and its
  material builds up a hot envelope around
  the remaining white dwarf. \citet{loren2009a} also find that when
  material of the secondary first hits the primary, the temperature at
  the interface becomes high enough to start carbon fusion. In their
  simulations, this leads to an rapid increase of the temperature
  there due to nuclear energy release. As the temperature increases,
  the material becomes non-degenerate, expands and cools down, thereby
  quenching the nuclear reactions. In the end, only a very small
  amount of material is burned.

  There is, however, an important difference in the
  interpretation. \citet{loren2009a} do not consider the possibility of
  a detonation forming at the hot-spot. In this case the detonation
  will be able to burn the whole white dwarf. This is a
  direct consequence of the fact that the detonation propagates
  supersonically and therefore burns all the material before it can
  expand.

  Hence, the crucial question that decides on the fate of a violent merger
  event is whether the merging process leads to the formation
  of a detonation or not. Unfortunately, the conditions and the process leading
  to the formation of a detonation are still not completely understood, but
  a topic of active research \citep[e.g.][]{seitenzahl2009b,roepke2007a}.

  \begin{figure*}[!t]
     \centering
     \includegraphics[width=0.9\linewidth]{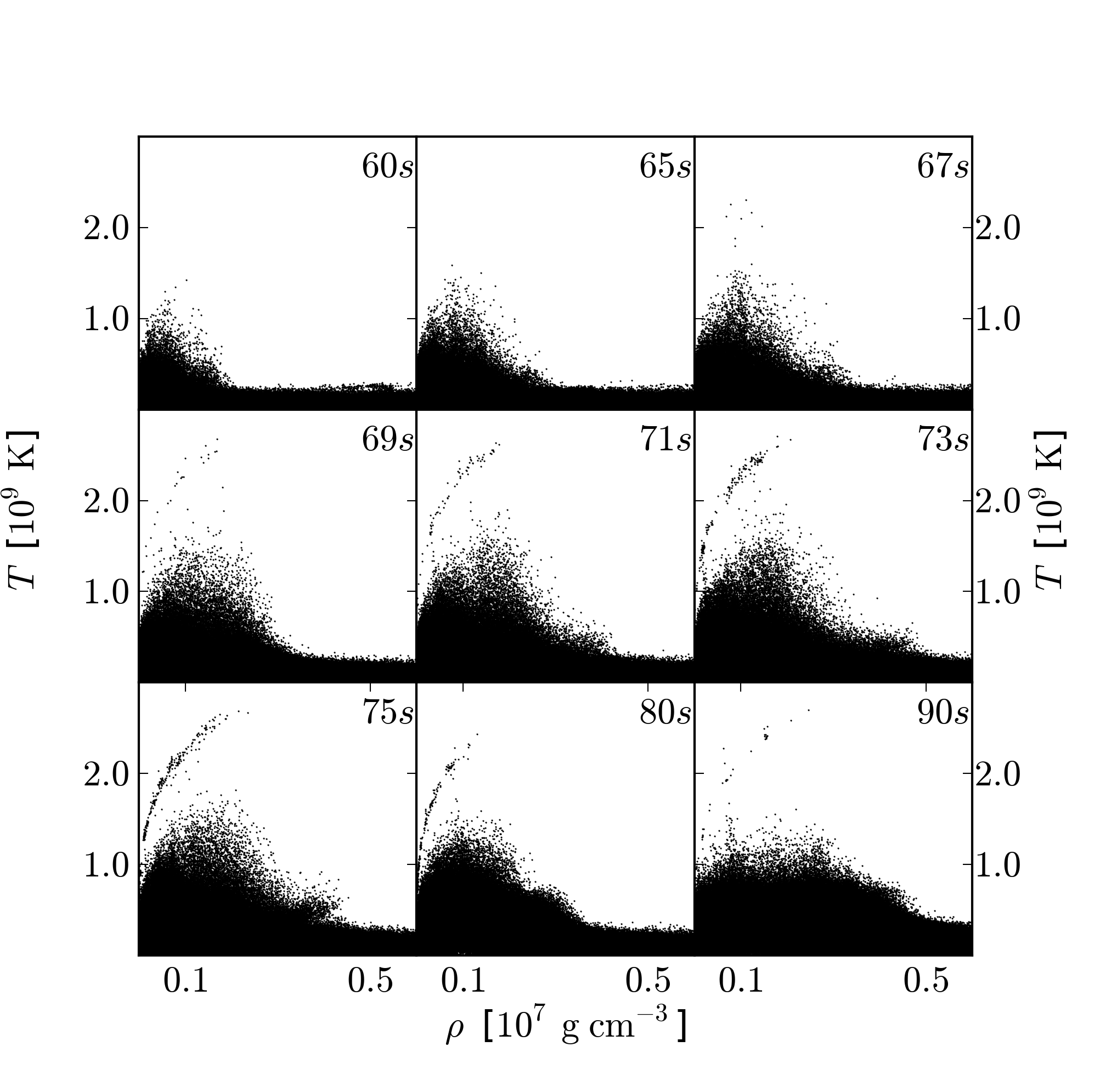}
     \caption{Density vs. temperature scatter plot for all particles of the merger of
       a $0.90 \, \mathrm{M}_{\odot}$ and a $0.81 \, \mathrm{M}_{\odot}$ white dwarf.} 
     \label{fig:rhotemp}
  \end{figure*}

  \subsection{Detonation ignition mechanisms}
  
  There are basically two ways a detonation can trigger:
  Firstly, a strong enough shock may lead to \emph{direct formation} of
  a detonation.  In this case, an already existing shock wave
  compresses and heats the material causing nuclear burning behind the
  shock. If the nuclear energy release behind the shock becomes large
  enough, it may be sufficient to create a self-sustained detonation
  \citep[e.g.][]{body1997a}. As the merger is very violent, we
  naturally expect strong shocks to occur in the interaction
  region. Unfortunately, however, we are far from being able to
  resolve these shocks on scales small enough to observe the formation
  of a detonation. As all SPH particles in our simulation have the
  same mass, the mass resolution is constant. Therefore regions with
  higher density are resolved better than regions with low
  density. Given that the interaction between the two white dwarfs
  occurs at comparatively low densities, their spatial resolution
  in the interaction region is intrinsically worse than at the center
  of the white dwarfs. In fact, we only barely resolve shocks in the
  interaction region at all. Therefore we are not able to judge from
  our current simulations whether a detonation can form directly from
  a shock. Since typical detonation flames are of the order of a few
  meters in width and the best resolved regions in our simulations are of the
  order of $10^{5} \mathrm{m}$, only future simulations with very high
  adaptivity may be able to resolve their formation in global simulations.

  Secondly, it may be possible to form a detonation spontaneously
  without a preceding shock.  In this case, as first proposed by
  \citet{zeldovich1970a}, nuclear burning in a preconditioned region
  (i.e.~with a spatial temperature or composition gradient that
  causes an induction time gradient)
  creates a spontaneous ignition wave. If the preconditioning is
  favorable, and this ignition wave has a supersonic phase
  velocity, it forms a shock wave that is followed by nuclear
  burning. But only if the release of nuclear energy behind the shock wave is
  sufficient to sustain the shock after it reaches the bottom of the
  gradient, it becomes a self-sustained detonation.

  \begin{figure*}[!ht]
     \centering
     \includegraphics[width=0.9\textwidth]{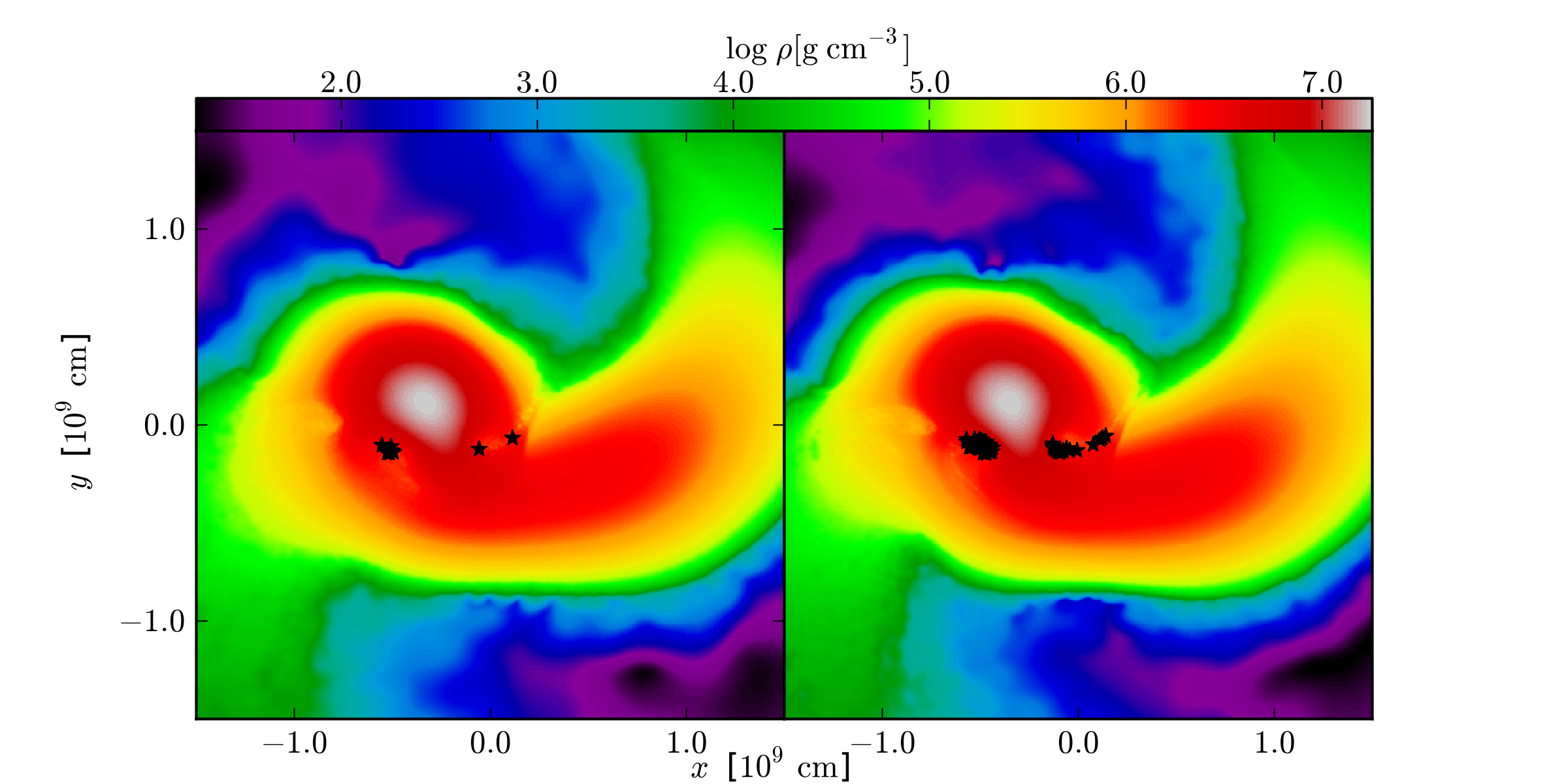}
     \caption{Resolution test of the detonation conditions of the
       merger of two white dwarfs of $0.90 \, \mathrm{M}_{\odot}$ and
       $0.89 \, \mathrm{M}_{\odot}$ with an initial period of $25
       \mathrm{s}$.  The left panel shows a simulation with $2 \cdot
       10^6$ particles, the simulation shown in the right panel
       contains $10^7$ particles. Both panels show a density slice
       after the system has evolved for $32 \mathrm{s}$. The black
       stars mark all particles with temperatures above $2 \cdot 10^9 \mathrm{K}$.}
     \label{fig:0989comparison}
  \end{figure*}
  
  Because the mechanism itself can not be resolved in our simulations,
  we have to rely on studies that model the mechanism
  for different conditions in a microscopic region. Such studies provide an indication whether
  a detonation is expected to form for the physical conditions in our simulation. In a recent
  study, \citet{seitenzahl2009b} carried out resolved one-dimensional
  simulations of the gradient mechanism for different initial
  conditions (i.e.~varying temperature, density and steepness of the
  gradient). They show that for the
  most favorable conditions of geometry and gradient the density
  threshold to form a detonation in carbon-oxygen material is
  $10^6\mathrm{g\ cm^{-3}}$. At this density, a temperature above
  $2.8 \cdot 10^9\mathrm{K}$ is required over a range of some
  meters. For higher densities, the temperatures required drop. In the
  following, we require a hot-spot with a density larger
  than $2 \cdot 10^6\mathrm{g\ cm^{-3}}$ and a temperature above
  $2.5 \cdot 10^9\mathrm{K}$ for reaching the conclusion that a detonation forms.

  \subsection{Conditions reached in the merger simulation}
  
  Figure~\ref{fig:rhotemp} shows density versus temperature of all
  particles for several snapshots from $60\,\mathrm{s}$ to
  $90\,\mathrm{s}$ after the start of the simulation in a scatter
  plot. This is the time-range when the secondary white dwarf is disrupted,
  its material is violently accreted onto the primary white dwarf and
  a hot-spot at the interface emerges. Afterwards, the temperature drops
  again as the hot material expands. Temperatures exceeding $2 \cdot 10^9 \, \mathrm{K}$
  are reached, but only for a small number of particles.
  These particles are the result of the additional energy release by thermonuclear
  combustion, because they ignite carbon which is subsequently burned. In simulations using
  identical initial conditions but no nuclear network they do not exist.
  Some of them are also in dense enough regions to fulfill the criterion
  we require to form a detonation.
  
  As the overall number of these particles is quite small,
  their exact properties have to be taken with caution. Therefore we
  have to check whether they show a barely resolved physical effect or
  are just a numerical artifact. This can be tested by redoing the
  simulation with significantly higher resolution.

  The binary is not completely merged at the time the first
  hot spots form. Therefore, if a detonation occurs, it stops any
  further evolution of the merger and the merged object never
  reaches the final state of a white dwarf surrounded by a hot
  envelope found in previous simulations
  \citep[e.g.][]{guerrero2004a, yoon2007a, loren2009a}.

  \subsection{Convergence study}
  
  For a convergence study,  we choose a slightly different binary with masses
  of $0.90\, \mathrm{M}_{\odot}$ and $0.89 \, \mathrm{M}_{\odot}$ for
  the two WDs, respectively. As this binary has a mass ratio
  much closer to one, it is expected to merge more violently and therefore it should
  be easier to find extreme conditions in it.
  
  Figure~\ref{fig:0989comparison} shows the density of two simulations
  of such a merger with $2 \cdot 10^6$ and $10^7$ particles,
  respectively. It demonstrates
  excellent agreement in the density structure between both
  simulations. In both simulations the particles with temperatures exceeding
  $2 \cdot 10^9 \mathrm{K}$ are found at the same locations. The number of
  these particles, however, increases from $11$ to $127$ from the
  lower resolved to the higher resolved simulation. As the number of
  particles is only increased by a factor of five, the total mass at
  temperatures above $2 \cdot 10^9 \mathrm{K}$ is approximately
  twice as large in the high resolution simulation. Therefore we can
  conclude that these hot particles are real, rather than a numerical
  artifact. The high resolution simulation indicates that we
  underestimate the conditions in these hot-spots due to a lack of
  resolution in our standard setup.

 \begin{figure*}[!ht]
     \centering
     \includegraphics[width=0.8\textwidth]{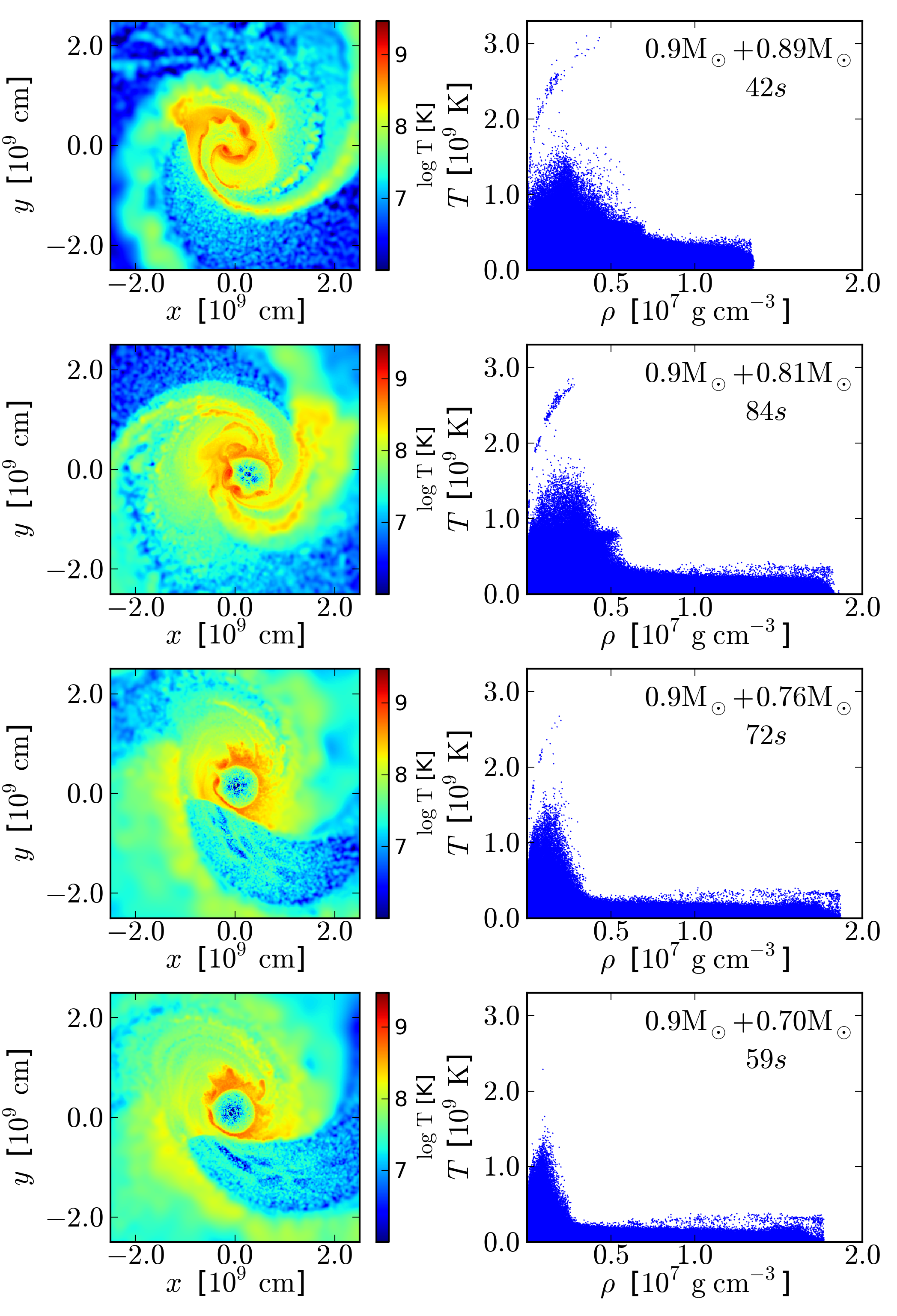}
     \caption[Mass ratio]
     {Temperature slice and density-temperature scatter diagram for
       four mergers with the same primary white dwarf, but different
       secondary white dwarfs, that have mass ratios of $0.99$, $0.9$, $0.84$
       and $0.78$. The rows show a temperature slice (on
       the left) and the distribution of all particles in
       temperature-density space (on the right) approximately at the time when there
       are the most favorable conditions for a
       detonation. Red/blue colors indicate high/low
       temperatures.}
     \label{fig:qtest09}
  \end{figure*}

  \subsection{Additional effects on the detonation conditions}

  There are two additional effects that may influence the formation of
  a detonation, but are basically ignored in our simulations.

  As discussed above, we do not resolve shocks at the interface very
  well. This effect can only be overcome by choosing a scheme that
  allows for refinement according to arbitrary criteria.

  Second, according to stellar evolution calculations there is a small
  helium shell on top of a carbon-oxygen white dwarf. For an isolated white
  dwarf around $0.9 \, \mathrm{M}_{\odot}$ this shell is expected to
  contain $\approx 10^{-3}$ to $10^{-2} \, \mathrm{M}_{\odot}$ of helium
  \citep{iben1985a}. It is known from the latest
  studies by \citet{seitenzahl2009b} that already a small mass
  fraction of helium mixed into carbon fuel significantly lowers the
  density required for the formation of a detonation.  As the expected
  helium layer is located exactly in the region where the merger is
  most violent and the hot-spots form, it may facilitate
  a detonation at even lower densities in our scenario.

  In summary, there are reasons to believe that detonations may form
  for even a wider range of conditions than the ones we have assumed here.

  \begin{figure*}[!t]
     \centering
     \includegraphics[width=0.75\textwidth]{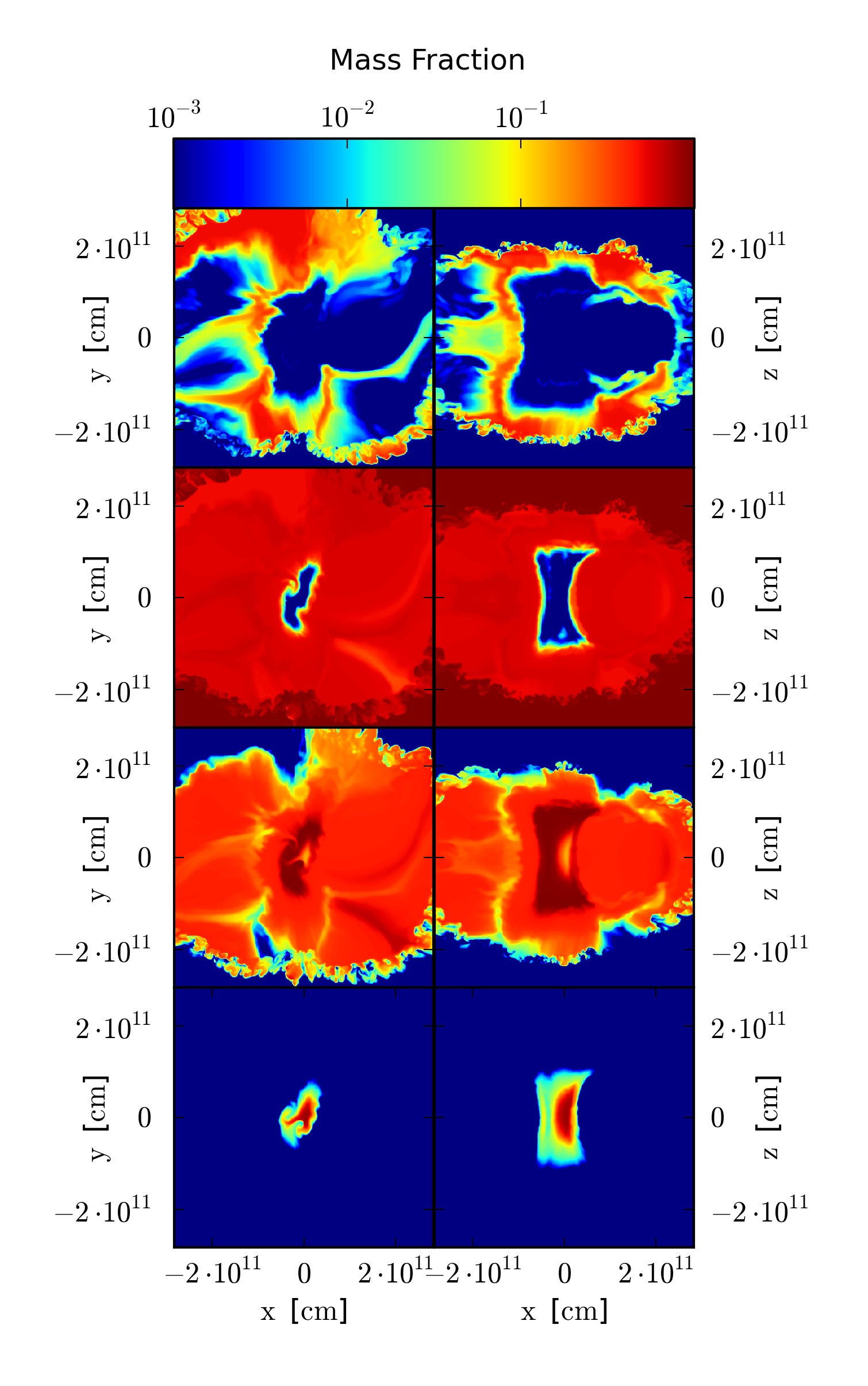}
     \caption{Final composition of the ejecta of the the explosion of a merger
       of two $0.89\, \mathrm{M_\odot}$ white dwarfs. At this time, $100 \mathrm{s}$
       after the explosion, the ejecta expand homologously. The panels show from the
       top to the bottom the abundances of carbon, oxygen,
       intermediate mass elements and iron group elements. The left
       and right panels show the system viewed face-on and edge-on,
       respectively.}
     \label{fig:comp}
  \end{figure*}

  \section{The impact of the mass ratio on the merger}
  \label{sec:mass_ratio}

  The violent merger scenario can only be realized frequently
  enough to account for a significant fraction of all type
  Ia supernovae if it also works for moderate mass differences
  between the two white dwarfs. Thus, we have to check how changing
  the mass ratio for binary systems with a fixed primary white dwarf mass
  affects the conditions during the merger. Only if mergers with a reasonable
  range also lead to the formation of a detonation, the scenario will be
  more than an exotic possibility. To test this, we compare
  four different mergers. All have the same primary
  mass of $0.9\, \mathrm{M_\odot}$, but different secondary masses of
  $0.89\, \mathrm{M_\odot}$, $0.81\, \mathrm{M_\odot}$, $0.76\, \mathrm{M_\odot}$ and $0.7\,
  \mathrm{M_\odot}$ which give mass ratios of $0.99$,
  $0.9$, $0.84$ and $0.78$.  In our simulations, they are set up with initial
  periods of $25\,\mathrm{s}$, $33\,\mathrm{s}$, $36\,\mathrm{s}$,
  and $40\,\mathrm{s}$, respectively.

  Figure~\ref{fig:qtest09} shows densities and temperatures of all
  particles of these simulations at the time when the conditions are
  most favorable for a detonation as well as temperature slices through the
  centers of the binaries. Obviously, there are considerable
  differences between these three systems. Firstly, with decreasing mass
  ratio, the merger becomes less violent. While the mergers with
  mass ratios of $0.99$, $0.9$ and $0.84$ produce several hot particles that
  ignite carbon and reach temperatures above $2 \cdot 10^9
  \mathrm{K}$, this is not the case for the merger with the smallest
  mass ratio. It is still possible that resolving the interaction region better
  will also show hotter particles. However, at
  the moment it seems more likely that below a certain mass ratio the
  merger is just not violent enough to ignite a detonation. This suggests
  a limiting mass ratio for the violent merger scenario of around $0.8$.
  
  Another difference is the dynamical effect of the merger on the primary
  white dwarf. In a nearly equal mass merger the primary star
  is heavily distorted, very similar to the equal mass merger described
  in \citet{pakmor2010a}. For a mass ratio of $0.9$, the primary white
  dwarf remains unaffected in the center, but its surface is
  distorted. For the smaller mass ratios, it stays completely
  intact and cool and is surrounded by the material of the
  disrupted less massive companion.

  The effect of the merger on the primary white dwarf can also be seen
  in the right panels of Figure~\ref{fig:qtest09}. In the $\mathrm{q} =
  0.99$ merger the central density of the remaining white dwarf is
  lower than in the other cases. As for these mergers the
  amount of $^{56}\mathrm{Ni}$ produced depends sensitively on the
  central density of the remaining white dwarf, this leads to smaller
  $^{56}\mathrm{Ni}$ masses in mergers with higher mass ratios.

  While this does not change the scenario fundamentally, it breaks the
  relation between mass of the primary white dwarf and the final
  $^{56}\mathrm{Ni}$ mass of the explosion for dim explosions. For
  explosions of mergers of more massive white dwarfs, however, where most of the
  $^{56}\mathrm{Ni}$ is produced in nuclear statistical equilibrium,
  this will be a minor effect only.  
  
  As shown in Figure~\ref{fig:qtest09},
  with decreasing mass ratio the merger becomes less violent and the
  density in the hot spots drops. This may make the formation of a detonation
  more difficult for smaller mass ratios.

  For the smallest mass ratio no violent merger occurs
  and we see the onset of a different merger regime in which an
  accretion disk forms around primary white dwarf.

  \section{Comparison with observations}
  \label{sec:observations}

  \subsection{Composition of the ejecta}

  \begin{figure}[!h]
     \centering
     \includegraphics[width=0.9\linewidth]{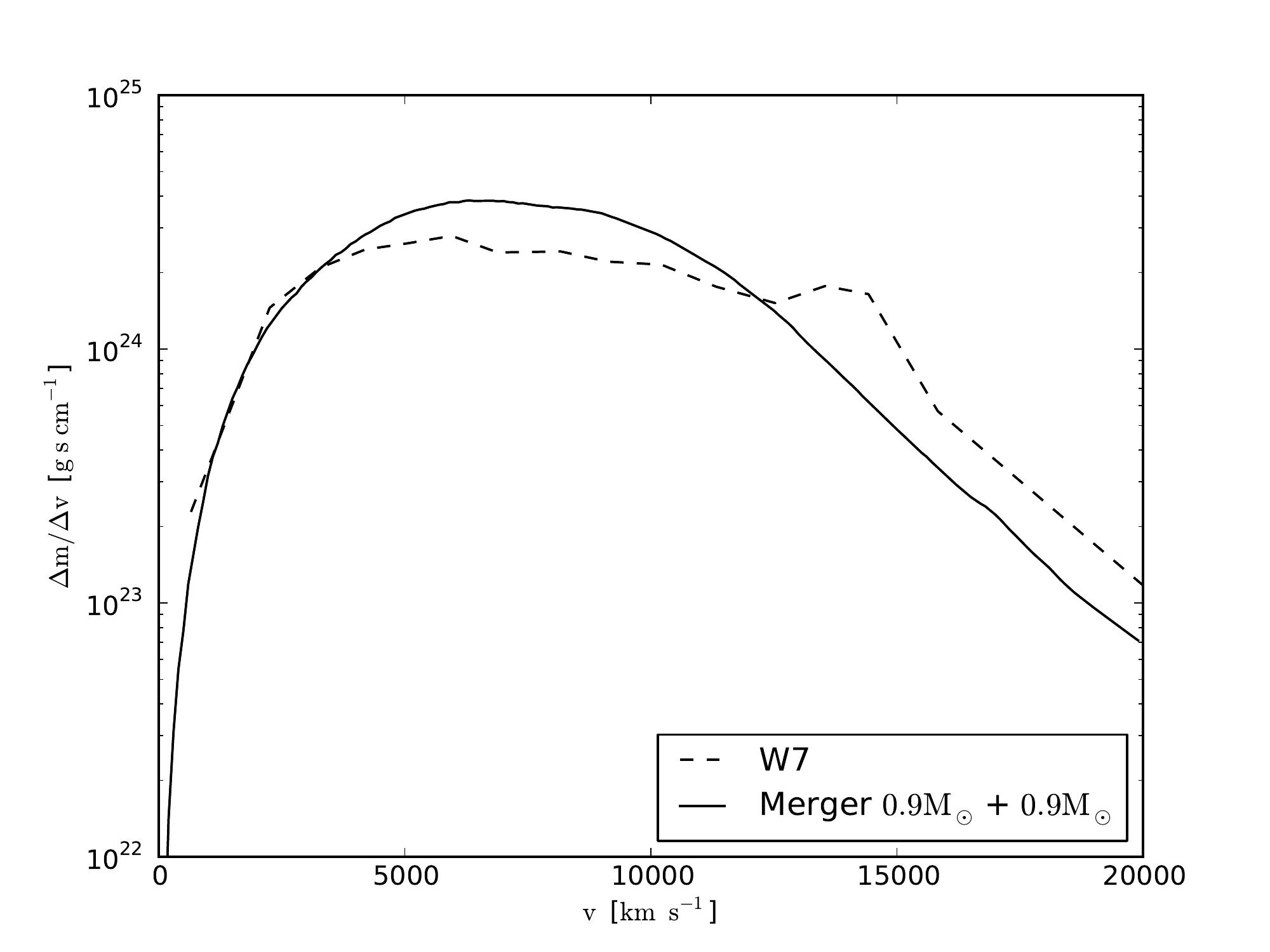}
     \caption{Mass distribution in velocity space resulting from the thermonuclear
              explosion following a violent merger of two white dwarfs
              of $0.9\, \mathrm{M_\odot}$. The standard Chandrasekhar-mass
              SN Ia model W7 is overplotted for comparison.}
     \label{fig:dmdv}
  \end{figure}
  
  \begin{figure}[!h]
     \centering
     \includegraphics[width=0.9\linewidth]{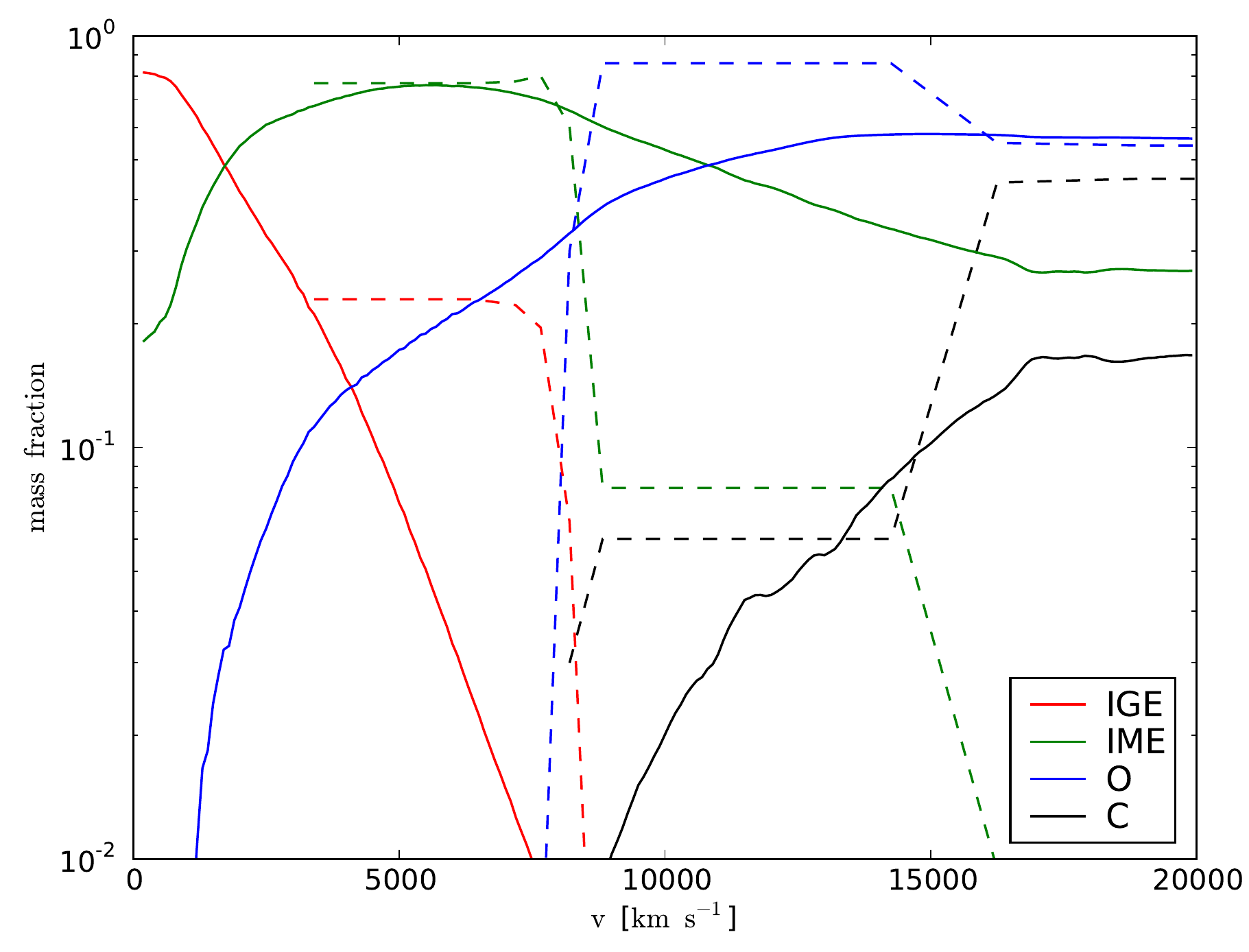}
     \caption{Angle averaged composition in velocity space (straight lines)
       of the result of the thermonuclear explosion following a violent merger
       of two white dwarfs of $0.9\, \mathrm{M_\odot}$. It is
       compared to the results of an abundance tomography study of SN
       2005bl by \citet{hachinger2009a} (dashed lines). Note that only a qualitative
       match can be expected, as the exact abundances depend on the
       density profiles, where \citet{hachinger2009a} used somewhat different
       assumptions.} 
     \label{fig:dxndv}
  \end{figure}

  Figure \ref{fig:comp} shows the ejecta composition of the thermonuclear
  explosion of the merger of two $0.89\, \mathrm{M_\odot}$ as described in
  \citet{pakmor2010a}, which is close to the merger with a mass ratio of $0.99$
  described above.
  
  The final composition contains $0.03 \, \mathrm{M_\odot}$ of carbon,
  $0.54 \, \mathrm{M_\odot}$ of oxygen, $1.05 \, \mathrm{M_\odot}$ of
  intermediate-mass elements and $0.1 \, \mathrm{M_\odot}$ of iron
  group elements. As the densities at the time the explosion happens
  do not exceed $2 \times 10^7 \mathrm{g\ cm^{-3}}$, nuclear burning
  does not reach nuclear statistical equilibrium anywhere in the object. Instead,
  iron group elements are only produced by incomplete silicon burning.

  As the nuclear burning in the detonation takes place
  at low densities, electron captures are not efficient. Therefore the
  initial electron fraction $Y_e$ of the unburned material is
  conserved throughout the nuclear burning. In case of $Y_e = 0.5$ the
  iron group elements consist of almost pure $^{56}\mathrm{Ni}$.
  A lower initial $Y_e$ is equivalent to more neutron-rich material in
  the pre-explosion composition (i.e.~more $^{22}$Ne resulting from the
  He burning phase). It leads to
  the production of some amount of stable iron replacing some of the
  $^{56}\mathrm{Ni}$. In total, however, the fraction of stable iron group
  elements produced is very small.

  As shown in Figure~\ref{fig:comp}, iron group elements are located at the
  center only, surrounded by intermediate-mass elements which are mixed with
  oxygen. Most carbon is found in the outermost parts, but there is
  some mixing with intermediate-mass elements. It is important to note
  that both, carbon and oxygen, are at different locations than the
  iron group elements, i.e.~there is no mixing of unburned material
  with iron group elements. There is also a clear difference between
  the distribution in the \textit{x}-\textit{y}-plane, which is the
  plane of rotation, and the distribution along the \textit{z}-axis.
  Along the \textit{z}-axis, the iron group elements are spread out
  much more, which means that in this direction there are iron group elements at
  higher velocities than within the \textit{x}-\textit{y}-plane. The opposite
  trend is observed for oxygen and intermediate mass elements, which
  reach further out in the plane of rotation than perpendicular to
  it. This is a result of the density structure of the merged object
  through which the detonation propagates.
  Consequently, lightcurves and spectra of this explosion are
  expected to show considerable viewing angle dependence \citep[for viewing
  angle dependent lightcurves see][]{pakmor2010a}.

  \subsection{Comparison with other models and observations}
  
  Compared to the standard W7 explosion model of \citet{nomoto1984a}
  our explosion models contains about 25\% more mass, but the kinetic
  energy of the ejecta is about the same. Therefore the velocities are lower on average.
  Figure~\ref{fig:dmdv} shows the mass distribution in velocity space
  in comparison to the W7 model. The merger has more mass at velocities
  below $12,000 \mathrm{km\ s^{-1}}$ and significantly less
  mass at higher velocities. With a $^{56}\mathrm{Ni}$ yield of only
  around $0.1 \, \mathrm{M_\odot}$ it is in the range of
  subluminous SNe~Ia. We argue that it falls into
  the class of 1991bg-like supernovae described by
  \citet[e.g.]{leibundgut1993a,taubenberger2008a}. For one 1991bg-like supernova, SN~2005bl,
  a detailed spectral analysis was presented by \citet{hachinger2009a}. In this
  study the authors find that decent models for the observed spectra
  require significantly less mass at higher velocities than what is inferred from the
  W7 mass profile. This is in agreement with the results of our
  simulations. Figure~\ref{fig:dmdv} shows a comparison between our
  final abundance pattern and the results of the abundance tomography
  for SN~2005bl. Despite differences in details, the main features are
  in qualitative agreement. There are basically three layers that
  differ in composition: The outermost layers are dominated by oxygen, with
  contributions of unburned carbon and some amount of intermediate
  mass elements. Below about $11,000 \, \mathrm{km\ s^{-1}}$ intermediate
  mass elements dominate down to very low velocities. Carbon is found down to $\approx 9,000 \,
  \mathrm{km\ s^{-1}}$. Iron group elements are present up to velocities
  of $\approx 7,500 \, \mathrm{km\ s^{-1}}$. In our model they dominate only in the
  very innermost part of the ejecta that is not probed by the photospheric spectra used
  in \citet{hachinger2009a}.
 
  The composition pattern of the ejecta of SN~2005bl inferred from abundance tomography is
  suggestive of a detonation at low densities as provided by our
  model. In particular, the fact that iron group elements co-exist with
  intermediate-mass elements down to very low velocities
  but -- at least as far as the observations can probe it --
  never dominate would be hard to explain with mixing alone. It
  strongly points to an extended range of incomplete silicon
  burning.

  Note that the abundances from our simulations shown in Fig.~\ref{fig:dmdv}
  are angle averaged. Taking into account the asymmetry of the
  explosion as shown in Fig \ref{fig:comp}, there is obviously
  some angle dependence in the composition in velocity space.
  This asymmetry will also affect the polarization of the observed
  spectra.  Even though no spectrapolarimetry data is available
  from SN~1991bg or SN~2005bl there exist polarization spectra
  from SN~1999by, which falls into the same class of
  objects. \citet{howell2001a} find that this supernova shows
  continuum polarization of $\approx 0.3\%-0.8\%$. This marks an
  important difference to normal SNe Ia, which do not show continuum
  polarization \citep{wang2008a}. Our model provides a natural explanation for the
  comparatively high degree of polarization in spectra of 1991bg-like objects,
  as the explosion is intrinsically asymmetric.

  As this model assumes a progenitor system that is most likely different from
  that of normal
  SNe Ia, it may also be able to explain the observed trend of 1991bg-like
  SNe to favor early-type galaxies and therefore old stellar
  populations \citep[e.g.][]{howell2001b}.
  A long inspiral time of the original WD-WD binary leads to a
  long delay time between the formation of the progenitor system and
  the supernova explosion in our model.

  Overall, the violent merger model presented here is not only the first realistic
  hydrodynamical model of a thermonuclear explosion of a WD that is
  able to produce $^{56}\mathrm{Ni}$ masses small enough to explain
  subluminous SN Ia, but it also shows remarkable agreement with observable
  features of 1991bg-like objects in a more
  detailed comparison. A more quantitative comparison based on
  synthetic lightcurves and spectra can be found in
  \citet{pakmor2010a}.

  \section{Conclusion}
  \label{sec:conclusion}

  In this paper we investigate the effect of variations in the mass
  ratio of merging white dwarf binaries on the \emph{violent merger scenario} proposed
  by \citet{pakmor2010a}.
  We first concentrate on a prototypical realization of this scenario: the
  case of a double degenerate binary system with a total mass of $1.61
  \, \mathrm{M_\odot}$ and a mass ratio of $0.9$. We follow the evolution of the binary
   system and subsequent merging event in smoothed particle hydrodynamical simulations.

  We discuss in detail the conditions that arise in the violent merging event
  and find that it is reasonable to assume that they lead to the formation of
  a detonation. We also show that while we are only barely resolving the
  most important part of the simulation (i.e.~the interface between the two
  white dwarfs, where the detonation forms), a very high resolution run
  comprising $10^7$ particles turns out to be even more in favor
  of the formation of a detonation.

  To estimate the influence on the mass ratio of the merging WDs on the detonation ignition
  conditions, we compare mergers with a primary white dwarf mass of 
  $0.9 \, \mathrm{M_\odot}$, but different mass ratios ranging from $0.76$
  to unity. We find that merging systems with a mass ratio above $\sim 0.8$ detonate,
  but those with a mass ratio below $0.8$ fail to do so.

  Finally, we compare the final abundance structure in the ejecta of a thermonuclear explosion
  after the violent merger of two white dwarfs with $\sim 0.9\, \mathrm{M_\odot}$ with
  observed properties of faint SN 1991bg-like supernovae.
  The ejecta have higher mass, but smaller velocities
  compared to the standard Chandrasekhar-mass W7 model. The abundance
  distribution of the model provides a good qualitative match with
  the composition found from the abundance tomography of SN~2005bl
  by \citet{hachinger2009a}. Moreover, the model could provide a natural explanation for the
  polarization found in SN~1999by and connect the preference of SN~1991bg-like objects to
  occur in old stellar populations with their progenitors.

  Of course, future studies are needed to explore the parameter space of
  different white dwarf masses and mass ratios in the violent merger scenario. 
  Only this will allow us to understand the complete range of possible outcomes
  and the contribution of this channel to the observed sample of SNe~Ia.

  \begin{acknowledgements}
    This work was supported by the Deutsche Forschungs\-gemeinschaft via
    the Transregional Collaborative Research Center TRR 33 ``The Dark
    Universe'', the Excellence Cluster EXC153 ``Origin and Structure
    of the Universe'' and the Emmy Noether Program (RO 3676/1-1). The simulations
    were carried out at the Computer Center of the Max Planck Society, Garching,
    Germany and the John von Neumann Institute for Computing (NIC) in J\"{u}lich, Germany
    (project HMU14).
  \end{acknowledgements}

  \bibliographystyle{aa}
  \bibliography{../../bibliography/bib/astrofritz.bib}

\end{document}